\documentclass[10pt,a4paper]{article}

\usepackage{amsmath}
\usepackage{amssymb}

\usepackage[top=2cm,bottom=2cm,left=2cm,right=2cm]{geometry}
\usepackage{graphicx}

\usepackage{cite}

\usepackage{color,ulem}
\usepackage{nameref}

\usepackage[labelfont=bf,labelsep=period,justification=raggedright]{caption}

\makeatletter
\renewcommand{\@biblabel}[1]{\quad#1.}
\makeatother

\date{}

\usepackage{color}
\definecolor{greencolor}{rgb}{0,0.5,0.2}
\definecolor{redcolor}{rgb}{.7,0.,0.}
\definecolor{bluecolor}{rgb}{0,0.,1.}
\definecolor{greycolor}{rgb}{.5,.5,.5}

\begin{document}

% Title must be 150 characters or less
\begin{flushleft}
{\Large
\textbf{Predictability of extreme events in social media}
}
% Insert Author names, affiliations and corresponding author email.
\\
Jos\'e M. Miotto$^{1,\ast}$, 
Eduardo G. Altmann$^{1}$
\\
\bf{1} Max Planck Institute for the Physics of Complex Systems, 01187 Dresden, 
Germany
\\
$\ast$ E-mail: jmiotto@pks.mpg.de
\end{flushleft}

% Please keep the abstract between 250 and 300 words
\section*{Abstract}
It is part of our daily social-media experience that seemingly ordinary items 
(videos, news, publications, etc.) unexpectedly gain an enormous amount of 
attention. 
Here we investigate how unexpected these extreme events are. 
We propose a method that, given some information on the items, quantifies the 
predictability of events, i.e., the potential of identifying in advance the 
most 
successful items.
Applying this method to different data, ranging from views in YouTube videos to 
posts in Usenet discussion groups, we invariantly find that the predictability 
increases for the most extreme events. 
This indicates that, despite the inherently stochastic collective dynamics of 
users, efficient prediction is possible for the most successful items. 

% Please keep the Author Summary between 150 and 200 words
% Use first person. PLoS ONE authors please skip this step. 
% Author Summary not valid for PLoS ONE submissions.   
% \section*{Author Summary}

\section*{Introduction}

When items produced in social media are abundant, the public attention is the 
scarce factor for which they compete~\cite{Simon,Huberman2007,wu2009feedback}.
Success in such {\it economy of attention} is very uneven: the distribution
of attention across different items typically shows heavy tails which resemble 
Pareto's distribution of income~\cite{pareto} and,
more generally, are an outcome of complex collective 
dynamics~\cite{PandBak,price1976general,Salganik2006,
stringer2010statistical, 
weng2012competition, onnela2010spontaneous,Ratkiewicz2010,
perc2014matthew}
and non-trivial maximizations of entropic 
functions~\cite{peterson2013maximum,Marsili}.
Increasing availability of large databases confirm the universality of these 
observations and renew the interest on understanding the dynamics of attention, 
see Tab.~\ref{tab.1}.

\begin{table}[!ht]
  \caption{{\bf Examples in which fat-tailed distributions of popularity across 
items have been reported.}}
  \begin{tabular}{l|l|l|l}
    \hline
    \hline
    System & Item & Attention measure & Refs. \\
    \hline
    Online Videos & video & views, likes & \cite{Crane2008} \\
    Discussion Groups & threads & posts, answers & \cite{Altmann2011}\\
    Publications & papers &citations, views & 
\cite{price1976general,stringer2010statistical,Wang04102013,PennerFortunato2013}
\\
    Twitter & tweet  &  retweets & \cite{weng2012competition} \\
    WWW & webpage & views & \cite{Ratkiewicz2010} \\
    Online Petitions & petition & signers & \cite{Yasseri2013}\\
    \hline
    \hline
  \end{tabular}
  \label{tab.1}
\end{table}

\begin{figure}[!ht]
\begin{center}
  \includegraphics{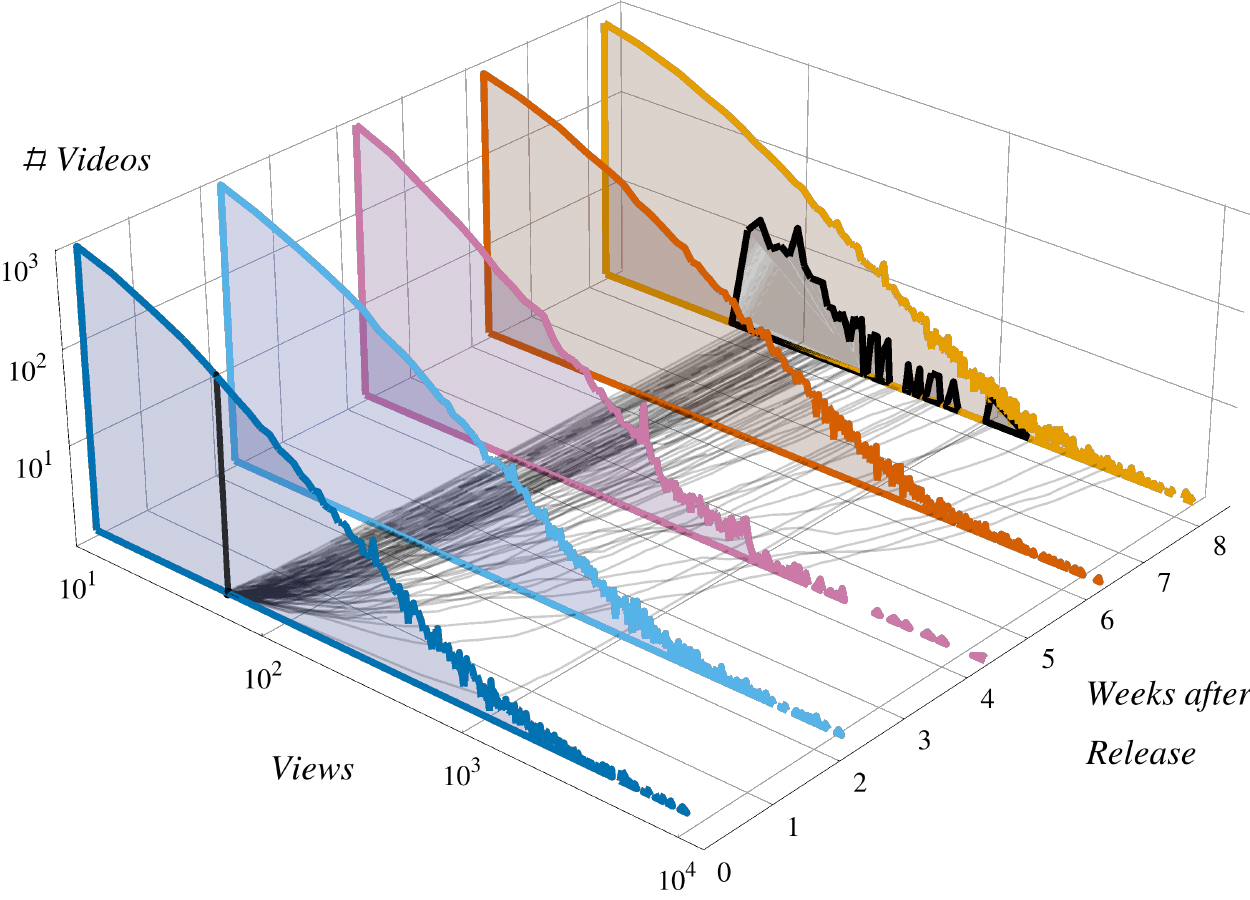}
\end{center}
\caption{
  {\bf Dynamics of views in YouTube.}
  {\bf Colored histograms:} distributions of views at fixed times after
  publication (0.3 million videos from our database).
  {\bf Gray lines at the bottom:}  trajectories of $120$ videos which had
    the same early success ($50$ views $2$ days after publication).
  {\bf Black histogram:} distribution of views of the $120$ selected videos $2$ 
months
  after publication.}
\label{fig.1}
\end{figure}

Universal features of heavy-tailed distributions do not easily lead to a good
forecast of specific items~\cite{PandBak}, a problem of major fundamental and 
practical interest~\cite{Wang04102013,bandari2012pulse,Sornette2004,Crane2008,
PennerFortunato2013}.
This is illustrated in Fig.~\ref{fig.1}, which shows that the heavy-tailed 
distribution appears at very short times but items with the same early success 
have radically different future evolutions.
The path of each item is sensitively dependent on idiosyncratic decisions which 
may be amplified through collective phenomena.

An important  question is how to quantify the extent into which prediction of 
individual items is possible (i.e., their {\it 
predictability})~\cite{KantzPredictionPredictability}.
Of particular interest --in social and natural systems-- is the predictability 
of extreme 
events~\cite{Albeverio2006extreme,hallerberg2007precursors,
hallerberg2008influence,Sornette2002,Ghil2011,Bogachev2011}, the small number 
of 
items in the tail of 
the distribution that gather a substantial portion of the public attention.

Measuring predictability is difficult because it is usually impossible to 
disentangle how multiple factors affect the quality of predictions.
For instance, predictions of the attention that individual items are going to 
receive rely on (i) information on properties of the item (e.g., metadata or 
the 
attention received in the first days) and (ii) a prediction strategy that 
converts the information into predictions.
The quality of the predictions reflect the interplay between these two factors 
and the dynamics of attention in the system.
In particular, the choice of the prediction strategy is crucial.
Instead, predictability is a property of the system and is by definition 
independent of the prediction strategy (it is the upper bound for the quality 
of any prediction based on the same information on the items).
A proper measure of the predictability should provide direct access to the 
properties of the system, enabling a quantification of the 
importance of different information on the items in terms of their predictive 
power.

In this paper we introduce a method to quantify the predictability of extreme 
events and apply it to data from social media.
This is done by formulating a simple prediction problem which allows for the
computation of the optimal prediction strategy.
The problem we consider is to provide a binary (yes/no) prediction whether an 
item will be an extreme event or not (attention passes a given threshold).
Predictability is then quantified as the quality of the optimal strategy.
We apply this method to four different systems: views of YouTube videos, 
comments in threads of Usenet discussion groups, votes to Stack-Overflow 
questions, and number of views of papers published in the journal PLOS ONE. 
Our most striking empirical finding is that in all cases the 
predictability increases for more extreme events (increasing threshold).
We show that this observation is  a direct consequence of differences in (the 
tails of) the distributions of attention conditioned by the known property
about the items.

The paper is divided as follows:
Sec.~\nameref{sec.attention} motivates the problem of event prediction 
by showing that it is robust to data with heavy tails.
Sec.~\nameref{sec.methods} introduces the method to quantify predictability, 
which is used in the Sec.~\nameref{sec.results}.
A summary of our findings appears in Sec.~\nameref{sec.conclusions}.

% TABLE 1

\section*{Motivation\label{sec.attention}}

\subsection*{Characterization of Heavy-tails}

Different systems in which competition for attention takes place share similar 
statistical properties.
Here we quantify attention of published items in 4 representative systems
(see Sec.~1 of the Supporting Information (SI) for details; all the data 
is available in Ref.~\cite{Miotto2014}):
\begin{itemize}
  \item views received by 16.2 million videos in YouTube.com between Jan. 2012 
    and Apr. 2013;
  \item posts written in 0.8 million threads in 9 different Usenet discussion 
    groups between 1994 and 2008;
  \item votes to 4.6 million questions published in Stack-Overflow between Jul. 
    2008 and Mar. 2013.
  \item views of 72246 papers published in the journal PLOS ONE from Dec. 2006 
    to Aug. 2013 (see also Ref.~\cite{fenner2013}).
\end{itemize}

The tails of the distribution $P(X)$ of attention $X$ (views, posts, etc.) 
received by the items (videos, threads, etc.) at a large time $t$ after 
publication is characterized without loss of generality using Extreme Value 
Theory.
It states that for large thresholds $x_p$ the probability $P(X|X>x_p)$ follows 
a 
Generalized Pareto distribution~\cite{coles2001introduction}

\begin{equation}\label{eq.powerlaw}
P(X>x|X>x_p) \sim \left(1+\frac{x-x_p}{\sigma\alpha}\right)^{-\alpha}.
\end{equation}
The fits of different partitions of our databases yield $\alpha \in [0.50,4.36]$
and are statistically significant already for relatively small $x_p$'s 
($p$-value$>0.05$ in $52$ out of $59$ fits, see SI Sec. 2 and Fig.~S1 for 
details).
These results confirm the presence of heavy tails, an observation reported
previously in a variety of cases (see Tab.~\ref{tab.1}). This suggests that our 
databases are representative of social media more 
generally (while scientific publications are usually not classified as social
media items, from the point of view of their online views, they are subject to 
the same attention-gathering process).

\subsection*{Prediction of Extreme Events}

Prediction in data with heavy tails is typically not robust.
As an example, consider using as a predictor $\hat{X}$ of the future attention
the mean $\hat{X}= \sum_{x=1}^\infty x P(x)$, which is the optimal predictor, 
if we measure the quality of prediction with the standard deviation of
$X$.
For heavy-tailed distributions, the mean and standard deviation may not be 
defined (for $\alpha<1$ and $\alpha<2$, respectively), making prediction not 
robust (i.e., it depends sensitively on the training and target datasets).
This illustrates the problems heavy-tails typically appear when value 
predictions are issued and indicates the need for a different approach to
prediction of attention.

We consider the problem of {\it event prediction} because, as shown below,
it is robust against fat-tailed distributions.
We say an event $E$ happens at time $t$ if the cumulative attention $X(t)$ 
received by the considered item until time $t$ is within a given range of 
values.
We are particularly interested in predicting extreme events $X(t)>x_*$,
i.e., to determine whether the attention to an item passes a threshold $x_*$ 
before time $t$.
The variable to be predicted for each item is binary: $E$ or $\bar{E}$ (not 
$E$).
We consider the problem of issuing binary predictions for each item ($E$ will 
occur or not), which is equivalent to a classification problem and different 
from a probabilistic prediction ($E$ will occur with a given probability).
Heavy tails do not affect the robusteness of the method because all items for 
which $X(t)>x_*$ count the same (each of them as one event), regardless of 
their 
size $x$.
Indeed, the tails of $P(X>x_*)$ determine simply how the probability of an 
event 
$P(E)$ depends on the threshold $x_*$ (we assume $P(X)$ exists).

%%%%%%%%%%%%%%%%%%%%%%%%%%%%%% Event prediction %%%%%%%%%%%%%%%%%%%%%%%%%%%%%%

% \section*{Predictability of Events\label{sec.event}}
\section*{Methods\label{sec.event}\label{sec.methods}}

In this section we introduce a method to quantify predictability based on
the binary prediction of extreme events.
This is done by arguing that, despite the seeming freedom to choose between 
different prediction strategies, it is possible to compute a single optimal 
strategy for this problem. We then show how the quality of prediction can be 
quantified and argue that the quality of the optimal strategy is a proper 
quantification of predictability.

Predictions are based on information on items which generally
lead to a partition of the items in groups $g\in\{1,\dots, 
G\}$ that have the same feature~\cite{sukhatme1994stratification}.
As a simple example of our general approach, consider the problem of 
predicting at publication time $t=0$ the YouTube videos that at $t=t_*=20$ 
days will have more than $x_*=1000$ views (about $P(E)\approx 6\%$ of all 
videos succeed).
As items' information, we use the category of a video so that, e.g., videos 
belonging to the category {\it music} correspond to one group $g$ and videos
belonging to {\it sport} correspond to a different group $g'$.
Since the membership to a group $g$ is the only thing that characterizes an 
item,
predictive strategies can only be based on the probability of having $E$ for 
that group, $P(E|g)$.

%\subsection*{Quality of a predicting strategy}

In principle, one can think about different strategies on how to issue binary 
predictions on the items of a group $g$. 
They can be based on the likelihood (L) $P(E|g)$ or on the posterior (P) 
probability~$P(g|E)$\cite{hallerberg2007precursors}, and they
can issue predictions stochastically (S), with rates proportional to the 
computed probabilities, or deterministically (D), only for the groups
with largest $P(g|E)$ or $P(E|g)$.
These simple considerations lead to four (out of many) alternative strategies 
to 
predict events (raise alarms) for items in group $g$
\begin{description}
\item[(LS)] stochastically based on the likelihood, i.e. with 
probability $\min\{1,\beta P(E|g)\}$, with $\beta\ge0$;
\item[(LD)] deterministically based on the likelihood, i.e. always 
if $P(E|g)>p_*$, with $0\le p_* \le 1$;
\item[(PS)] stochastically based on the posterior, i.e. with 
probability $\min\{1,\beta' P(g|E)\}$, with $\beta'\ge0$;
\item[(PD)] deterministically based on the posterior, i.e. always 
if $P(g|E)>p'_*$, with $0\le p'_* \le 1$.
\end{description}
In the limit of large number of predictions (items), the fraction of events 
that strategy (LS) predicts for each group $g$ matches the probability of 
events 
$P(E|g)$ and therefore strategy (LS) is 
{\it reliable}~\cite{brocker2009reliability} and can be considered a 
natural extension of a probabilistic predictor.
Predictions of strategies (LD), (PS) and (PD) do not follow $P(E|g)$ and 
therefore they are not reliable.

\begin{figure}[!ht]
\begin{center}
  \includegraphics{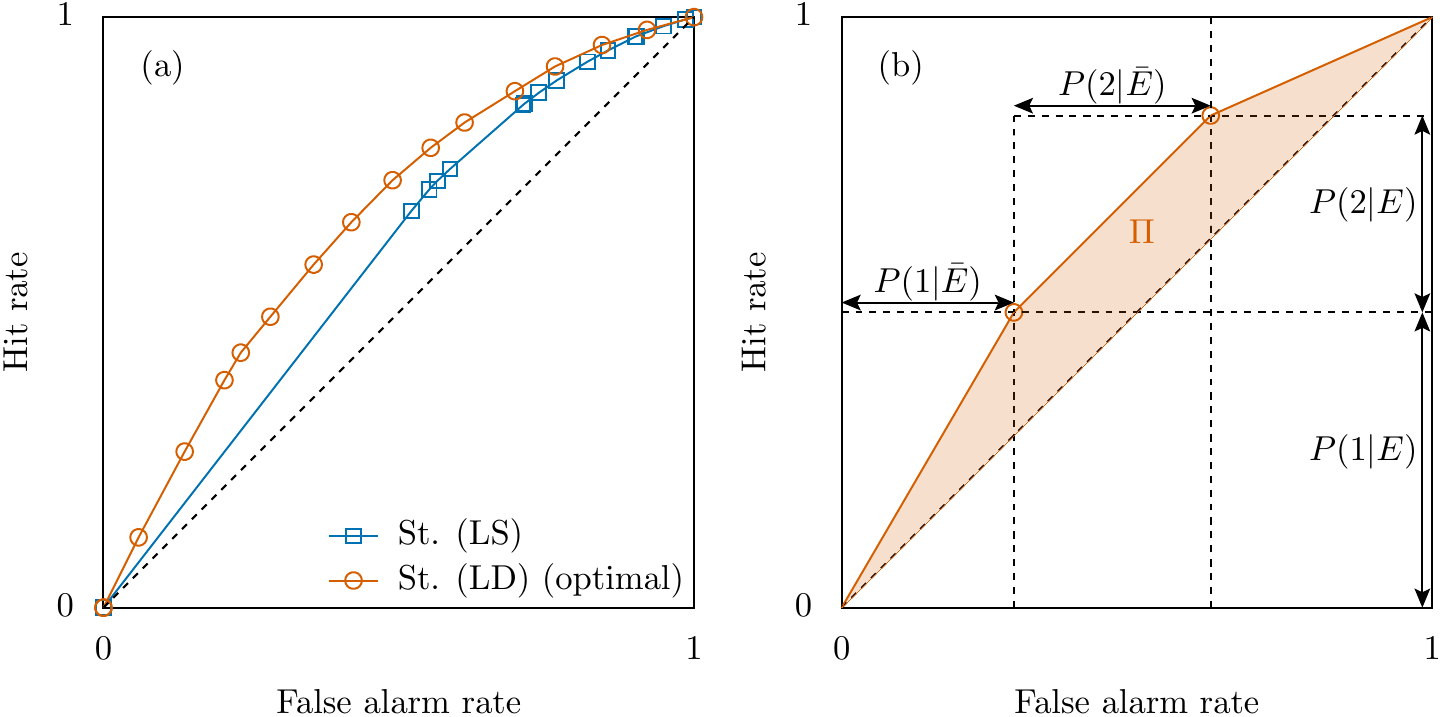}
\end{center}
\caption{
  {\bf Quantifying the quality of event-prediction strategies requires measuring
  both the hit and false alarm rates.}
  {\bf(a)} Performance of Strategy (LS) and Strategy (LD) for the problem of 
  predicting views of YouTube videos 20 days after publication based on their 
  categories.
  The symbols indicate where the rate of issued predictions for a given group 
  equals 1 (the straight lines between the symbols are obtained by issuing 
  predictions randomly with a growing rate).
  {\bf(b)} Illustration of the prediction curve (red line) for an optimal 
strategy 
  with three groups $g=1,2,3$ with $P(1)=P(2)=P(3)=1/3$ and $P(E|1)=0.3, P(E|
  2)=0.2, P(E|3)=0.1$.
}
\label{fig.2}
\end{figure}

The quality of a strategy for event prediction is assessed by computing the 
false alarm rate (or False Positive Rate, equal to one minus the specificity) 
and the hit rate (True Positive Rate, equal to the sensitivity) over all
predictions (items), see Appendix for details.
Varying the amount of desidered false alarms of the prediction strategy 
($\beta, p_*, \beta',$ and $ p_*'$ in the examples above), a curve in the 
hit$\times$false-alarm space is obtained, see Fig.~\ref{fig.2}(a).
The overall quality is measured by the area below this curve, known as
Area Under the Curve (AUC)~\cite{ROC1}. For convenience, we use the
area between the curve and the diagonal (hits=false-alarms), $\Pi=2\text{AUC}-1$
(equivalent to the Gini coefficient). In this way, $\Pi_S\in(-1,1)$ represents
the improvement of strategy $S$ against a random prediction.
In absence of information $\Pi_S=0$ and perfect predictions lead to $\Pi=1$.
In the YouTube example considered above, we obtain
$\Pi_{PS}<\Pi_{LS}<\Pi_{PD}<\Pi_{LD}$ (17\%, 18\%, 29\%, 32\%), 
indicating that strategy (LD) is the best one.

We now argue that strategy (LD) is optimal (or 
{\it dominant}~\cite{provost1998case}), i.e., for any false alarm rate it leads 
to a larger hit rate than any other strategy based on the same set of $P(E|g)$.
To see this, notice that strategy (LD) leads to a piecewise linear curve, see 
Fig.~\ref{fig.2}(b), and is the only ordering of the groups that enforces 
convexity in the hit$\times$false-alarms rates space, see 
Appendix~\ref{sec.dominant} for a formal derivation.
The ranking of the groups by $P(E|g)$ implies a ranking of the items, an 
implicit assumption in the measure of the performance of classification 
rules~\cite{ROC1,ROC2}.
The existence of an optimal strategy implies that the freedom in choosing the 
prediction strategy argued above is not genuine and that we can ignore the 
alternative strategies.
In our context, it implies that the performance of the optimal  strategy 
measures 
a property of the system (or problem), and not simply the efficiency of a 
particular strategy.
Therefore, we use the quality of prediction of the optimal strategy 
($\Pi\equiv\Pi_{LD}$) to quantify the predictability (i.e., the potential 
prediction) of the system for the given problem and information.
By geometrical arguments we obtain from Fig.~\ref{fig.2} (b) (see Appendix)
\begin{equation}\label{eq.pi}
\Pi=\sum_{g}\sum_{h<g}\dfrac{P(g)P(h)\left(P(E|h)-P(E|g)\right)}{P(E)(1-P(E))}, 
\end{equation}
where $P(g)$ is the probability of group $g$ and $g$ is ordered by decreasing 
$P(E|g)$, i.e., $h<g \Rightarrow P(E|h) > P(E|g)$.  

The value of $\Pi$ can be interpreted as the probability of
a correct classification of a pair of $E$ and $\bar{E}$ items~\cite{ROC1,ROC2}.
In practice, the optimality of this strategy is dependent on the estimation of 
the ordering of the groups according to $P(E|g)$. Wrong ordering may occur due 
to finite sampling on the training dataset or non-stationarities in the data. 
In fact, any permutation of indexes in Eq.~(\ref{eq.pi}) reduces $\Pi$.

% FIGURE 2

\section*{Results}

\subsection*{Application to Data\label{sec.results}}

Here we apply our methodology to the four social-media data described above.
We consider the problem of predicting at time $t_1\ge 0$ whether the attention 
$x$ of  an item at time $t_*>t_1$ will pass a threshold $x_*$. 
In practice, the calculation of $\Pi$ from the data is done counting the 
  number of items: (i) in each group $g$ [$P(g)=\allowbreak (\text{\# items in 
} 
g)/
  \allowbreak (\text{\# items})$]; (ii) that lead to an event 
  $[P(E)=\allowbreak (\text{\# items that crossed the threshold } \allowbreak 
x_* \text{ at } 
  \allowbreak t_*)/ \allowbreak (\text{\# items})]$; and  (iii) that lead to an 
  event given that they are in group $g$ $[P(E|g)=\allowbreak (\text{\# 
  items in }g$ \\$ \text{ that crossed the threshold }\allowbreak x_*\text{ at }
  t_*)/\allowbreak (\text{\# items in }g)]$. Finally, the groups are numbered 
as 
$g=1,2,
  \ldots, G$ by decreasing $P(E|g)$ and the sum over all groups is computed as 
  indicated in Eq.~(\ref{eq.pi}). In Ref.~\cite{Miotto2014} we provide
  a python script which performs this calculation in the data.

We report the values of $\Pi$ obtained from Eq.~(\ref{eq.pi}) considering two 
different informations on the items: 
\begin{itemize}
\item[1)] the attention at prediction time $x(t_1)$; 
\item[2)] information available at publication time $t=0$ (metadata).
\end{itemize}
In case 1), a group $g$ corresponds to items with the same $x(t_1)$.
These groups are naturally ordered in terms of $P(E|g)$ by the value of 
$x(t_1)$ and therefore the optimal strategy is equivalent to issue 
positive prediction to the items with $x(t_1)$ above a certain threshold.
In case 2),  the groups correspond to items having the same meta-data (e.g.,
belonging to the same category). In this case, we order the groups according 
to the empirically observed $P(E|g)$ (as discussed above).
Before performing a systematic exploration of parameters, we illustrate our 
approach in two examples :

\begin{itemize}

%\subsubsection*{YouTube}
\item Consider the case of predicting whether YouTube videos at $t_*=20$ days 
will have more than $x_*=1,000$ views. For case 1), we use the views achieved 
by 
the items after $t_1=3$ days and obtain a predictability of $\Pi=90\%$.
For case 2), we obtain that using the day of the week to group the items leads 
to $\Pi=3\%$ against $\Pi=31\%$ obtained using the categories of the videos.
This observation, which is robust against variations of $x_*$ and $t_*$, shows
that the category but not the day of the week is a relevant information in 
determining the occurrence of extreme events in YouTube.

%\subsubsection*{PLOS ONE}

\item Consider the problem of identifying in advance the papers published in 
the 
online journal PLOS ONE that received at least $7500$ views 2 years after 
publication, i.e $X(t_*=2 \text{years})>x_*=7500$ (only $P(E)=1\%$ achieve this 
threshold).
For case 1), knowing the number of views at $t_1=2$ months after publication
leads to a predictability of $\Pi=93\%$. For case 2), a predictability
$\Pi=19\%$ is achieved alone by knowing the number of authors  
of the paper --surprisingly, the chance of achieving a large number of views 
decays monotonously with number of author ($g$ increases with number of
authors).

\end{itemize}

\begin{figure}[!ht]
\begin{center}
\includegraphics{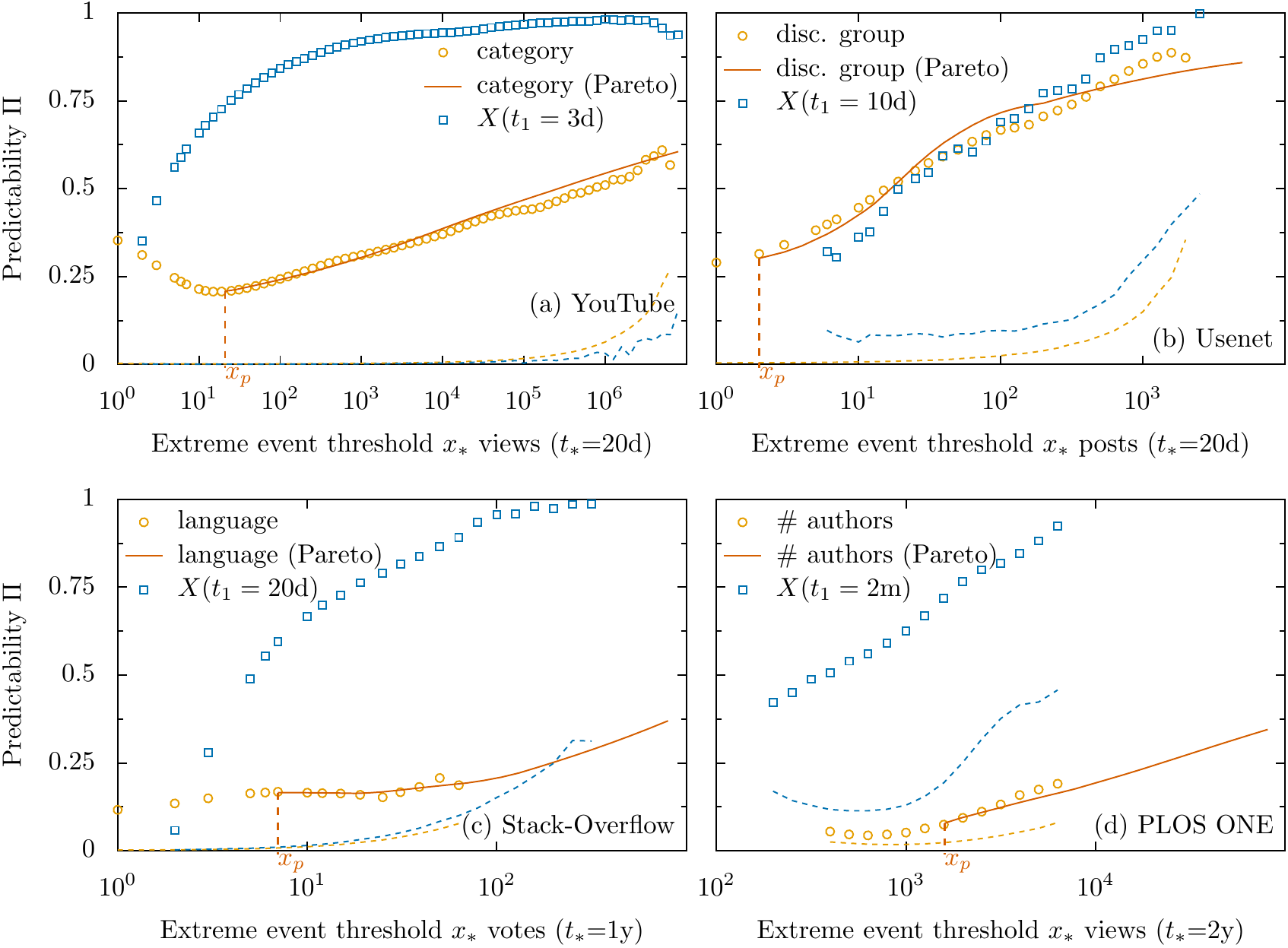}
\end{center}
\caption{
  {\bf Predictability increases for extreme events.}
  If the attention an item receives at time $t_*$ is above a threshold, 
  $X(t_*)>x_*$, an event $E$ is triggered.
  The plots show how the predictability $\Pi$ changes with $x_*$ using two 
  different informations to combine the items in groups $\{g\}$.
  {\bf Black circles:} $\Pi$ at time $t=0$ using metadata of the items to group 
  them.
  The {\bf red lines} are computed using as probabilities $P(E|g)$ the Extreme 
  Value distribution fits for each group at a threshold value $x_p$, see 
  Eq.~(\ref{eq.powerlaw}) and SI Sec. 2.
  {\bf Blue squares}: $\Pi$ at time $t_1<t_*$ using $X(t_1)$, i.e., the 
  attention the item obtained at day $t_1$.
  The {\bf dashed lines} are the values of the 95\% percentile of the
  distribution generated by measuring $\Pi$ in an ensemble of databases 
obtained 
  shuffling the attribution of groups ($g$) to items
  (the colors match the symbols and symbols are shown only where $\Pi$ 
  is at least twice this value).
  Results for the four databases are shown:
    {\bf(a)} YouTube 
      ($X$: views of a video; metadata: video category);
    {\bf(b)} Usenet discussion groups
      ($X$: posts in a thread; metadata: discussion group of the thread);
    {\bf(c)} Stack-Overflow
      ($X$: votes to a question; metadata: programming language of the 
question, 
        see SI Sec. 2 for details);
    {\bf(d)} PLOS ONE 
      ($X$: online views of a paper; metadata: number of authors of the paper).
  }
\label{fig.3}
\end{figure}

The examples above show that formula~(\ref{eq.pi}) allows for a quantification
of the importance of different factors (e.g., number of authors, 
early views to the paper) to the occurrence of extreme events, beyond
correlation and regression methods (see also Ref.~\cite{PennerFortunato2013}).
Besides the quantification of the predictability of specific problems, by
systematically varying $t_1, t_*,$ and $x_*$  we can quantify how the
predictability changes with time and with event magnitude.
Our most significant finding is that in all tested databases and grouping
strategies the predictability increases with $x_*$, i.e., extreme events become 
increasingly more predictable, as shown in Fig.~\ref{fig.3}.

%%%%%%%%%%%%%%%%%%%%%%%%%%%%%%% Extreme Events %%%%%%%%%%%%%%%%%%%%%%%%%%%%%%%

\subsection*{Discussion}

We now explain why predictability increases for extreme events (increasing 
$x_*$). 
We first show that this is not due to the reduction of the number of events
$P(E)$. Consider the case in which $E$ is defined in the interval
$[x_f-\Delta x,x_f+\Delta x)$. Assuming $P(X)$ to be smooth in $X$, for $\Delta 
x
\rightarrow 0$ at fixed $x_f$ we have that $P(E)\rightarrow P(x_f)\Delta x$ and
$P(E|g)\rightarrow P(x_f|g)\Delta x$ ($P(g)$ remains unaffected), and 
Eq.~(\ref{eq.pi})
yields
\begin{equation}  
\Pi=\frac{\sum_{g}\sum_{h>g}P(g)P(h)\left(P(x_f|h)-P(x_f|g)\right)}{P(E_f)[
1-\Delta x P(x_f)]},
  \label{eq.pi2}
\end{equation}
which decreases with $\Delta x\rightarrow 0$. This shows that the increased 
predictability with $x_*$
is not a trivial consequence of the reduction of $P(E)$ 
($\Delta x\rightarrow 0$), but
instead is a consequence of the change in $P(E|g)$ for extreme events $E$.

Systematic differences in the tails of $P(X|g)$ lead to an
increased predictability of extreme events.
Consider the case of two groups with cumulative distributions $P(E|g)$ that 
decay as a power law as in Eq.~(\ref{eq.powerlaw}) with exponents $\alpha$ and 
$\alpha'=\alpha+\epsilon$, with $P(1)=P(2)$.
From Eq.~(\ref{eq.pi}), $\Pi$ for large $x_*$ ($1-P(E)\approx1$) can be 
estimated as
\begin{equation}
\Pi = \frac{1}{4}\frac{P(E|1)-P(E|2)}{P(E|1)+P(E|2)} 
    = \frac{1}{4} \frac{x_*^{-\alpha}-x_*^{-(\alpha+\epsilon)}}
      {x_*^{-\alpha}+x_*^{-(\alpha+\epsilon)}}
    \approx \frac{1}{8}\log(x_*) \epsilon,
\end{equation}

where the approximation corresponds to the first order Taylor expansion around
$\epsilon=0$.
The calculation above can be directly applied to the results we obtained 
issuing predictions based on metadata.
The logarithmic dependency in Eq. (4) is consistent with the roughly linear 
behavior observed in Fig.~\ref{fig.3}(a,b).
A more accurate estimation is obtained using the power-law fits of
Eq.~(\ref{eq.powerlaw}) for each group $g$ and introducing the $P(E|g)$ 
obtained from these fits in Eq.~(\ref{eq.pi}). The red line in 
Fig.~\ref{fig.3} shows that this estimation agrees with the observations 
for values $x_* \gtrapprox x_p$, the threshold used in the fit.
Deviations observed for $x_* \gg x_p$ (e.g., for PLOS ONE data in panel (d)) 
reflect the deviations of $P(E|g)$ from the Pareto distribution obtained for 
small thresholds $x_p\ll x_*$.
This allows for an estimation of the predictability for large thresholds $x_*$ 
even in small datasets (when the sampling of $E$ is low).

A similar behavior is expected when prediction is performed based on the 
attention obtained at short times $t_1$. Eq.~(\ref{eq.pi2}) applies in this 
case 
too and therefore the increase in predictability is also due to change in 
$P(E|g)$ with $x_*$ for different $g$ (and not, e.g., due to the decrease of 
$P(E)$).
For increasingly large $x_*$ the items with significant probability of passing 
threshold concentrate on the large $x(t_1)$ and increase the predictability of 
the system.
We have verified that this happens already for simple multiplicative stochastic 
processes, such as the geometric Brownian motion (see Fig.~S2). 
This provides further support for the generality of our finding.
The dynamics of attention in specific systems affect the shape of 
predictability 
growth with threshold.

Altogether, we conclude that the difference in (the tails of) the distribution 
of attention of different groups $g$ is responsible for the increase in 
predictability for extreme events: for large $x_*$, any informative property on 
the items increases the relative difference among the $P(E|g)$. 
This corresponds to an increase of the information contained in the grouping
which leads to an increase in $\Pi$.

%%%%%%%%%%%%%%%%%%%%%%%%%%%%%%%% Conclusions %%%%%%%%%%%%%%%%%%%%%%%%%%%%%%%%

\section*{Conclusions}\label{sec.conclusions}

In summary, we propose a method, Eq.~(\ref{eq.pi}), to measure the 
predictability of extreme events for any given available information on the 
items. We applied this measure to four different social media databases and 
quantified how predictable the attention devoted to different items is and how 
informative are different properties of the items.
We quantified the predictability due to metadata available at publication 
date and due to the early success of the items and found that usually the 
latter quickly becomes more relevant than the former\footnote{Our results can
  also be applied for combinations of different informations on the items 
  (e.g., a group $g$ can be composed by videos in the category  {\it music} 
with 
  a fixed $x(t_1)$).
  In practice, the number of groups $G$ should be much smaller than the 
  observations in the training dataset to ensure an accurate estimation of 
  $P(E|g)$.}. 
Our most striking finding is that extreme events are better predictable than
non-extreme events, a result previously observed in physical
systems~\cite{hallerberg2008influence} and in time-series 
models~\cite{hallerberg2007precursors,Bogachev2011}.
For social media, this finding means that for the large attention catchers
the surprise is reduced and the possibilities to discriminate success enhanced. 

These results are particularly important in view of the widespread
observation of fat-tailed distributions of attention, which imply that 
extreme events carry a significant portion of the total public attention. 
Similar
 distributions appear in financial markets, in which case our methodology can 
quantify the
 increase in predictability due to the availability of specific information 
(e.g., in
 Ref.~\cite{preis2013quantifying} Internet activities were used as information 
to issue predictions).
For the numerous models of collective behavior leading to fat
tails~\cite{price1976general,stringer2010statistical,  
weng2012competition,onnela2010spontaneous,Ratkiewicz2010,PennerFortunato2013,
Wang04102013},
the predictability we estimate is a bound to the quality of binary
event predictions. 
Furthermore, our identifications of 
the factors leading to an improved predictability indicate which properties 
should be included in the models and which ones can be safely ignored 
(feature selection).
For instance, the relevant factors identified in our analysis should 
affect the
  growth rate of items in  rich-get-richer
  models~\cite{Ratkiewicz2010,perc2014matthew} or the transmission rates between 
agents in
  information-spreading models~\cite{RMP}.
The use of $\Pi$ to identify relevant factors goes beyond 
simple correlation tests and can be considered as a measure of causality in the 
sense of Granger~\cite{granger1980}.

Predictability in systems showing fat tails has been a matter of intense 
debate. While simple models of self-organized criticality suggest that 
prediction of individual events is impossible~\cite{PandBak}, the existence of 
predictable mechanisms for the very extreme events has been advocated in 
different systems~\cite{Sornette2002}.
In practice, predictability is not an yes/no 
question~\cite{Salganik2006,KantzPredictionPredictability} and the main 
contribution of this paper is to provide a robust quantification of the 
predictability of extreme events in systems showing fat-tailed distributions.

\section{Appendix}

\subsection{Quality of binary predictions}\label{sec.quality}

Comparing binary predictions and observations gives four possible results, 
given by the combination of the prediction (positive or negative) and its 
success (true or false).
If $A$ denotes the prediction of an event (an alarm), the hit rate (or True 
Positive Rate) and the false alarm rate (or False Positive Rate) are defined as
\begin{equation}\label{eq.A}
\begin{array}{ll}
  \text{hit rate} \equiv \dfrac{\text{number of true positives}}{\text{number of 
positives}}=
  P(A|E), \\ \\ \text{false alarm rate} \equiv \dfrac{\text{number of false 
positives}}{\text{number of
      negatives}} = P(A|\bar{E}).
\end{array}
\end{equation}
These are analogous to  measures like Accuracy and Specificity or Precision and 
Recall.
Prediction strategies typically have a specificity parameter (e.g., controlling 
the rate of false positives).
Varying this parameter, a prediction curve that goes from $(0,0)$ to $(1,1)$ is 
built in the hit$\times$false-alarm space.

\subsection{Demonstration that strategy LD (Bayes classifier) is 
dominant}\label{sec.dominant}

A strategy is dominant when for any given false alarm rate, the hit rate is 
maximized.
Following definition~(\ref{eq.A}), we write the $x$ and $y$ coordinates 
of the hit$\times$false-alarm plot as
\begin{equation} \label{eq.yx}
\begin{array}{ll}
  \text{hit rate} \equiv P(A|E) = \sum_{g=1}^G P(A|g) P(g|E) = \sum_{g=1}^G 
\pi_g y_g \equiv y,
  \\ \\
  \text{false-alarm rate} \equiv P(A|\bar{E}) = \sum_{g=1}^G P(A|g) P(g|\bar{E}) 
= \sum_{g=1}^G \pi_g
  x_g \equiv x,
\end{array}
\end{equation}
where for notational convenience $y_g\equiv P(g|E)$, $x_g\equiv
P(g|\bar{E})$, and $\pi_g\equiv P(A|g)$. Since predictions are issued 
based only on the information about the groups, strategies (both 
deterministic and stochastic) are defined uniquely by $\pi_g$, 
while $x_g$ and $y_g$ are estimated from data.
The computation of the dominant strategy corresponds to finding the 
$\pi_g$'s that maximize $y$ with the constraint $\sum_{g=1}^G \pi_g x_g=x$.
This problem can be solved exactly by applying the simplex method. % [CITE!].
Define $h$ such that $\sum_{g<h} x_g < x< \sum_{g\le h} x_g$;
we write Eq.~(\ref{eq.yx}) as:
\begin{equation}
\begin{aligned}
  y-\sum_{g<h} y_g  &= -\sum_{g<h}(1-\pi_g)y_g + \sum_{g>h} \pi_g y_g + \pi_{h} 
y_{h}, \\
  x-\sum_{g<h} x_g  &= -\sum_{g<h}(1-\pi_g)x_g + \sum_{g>h} \pi_g x_g + \pi_{h} 
x_{h}.
\end{aligned}
\end{equation}
Isolating $\pi_h$ in the lower equation and introducing it in the top one we 
obtain
\begin{align}\label{eq.y}
  y =& \sum_{g< h} y_g +x\frac{y_{h}}{x_{h}} \\
  & - \sum_{g<h}(1-\pi_g) x_g 
  \left( \frac{y_g}{x_g}-\frac{y_{h}}{x_{h}} \right) + 
  \sum_{g>h}\pi_g x_g \left( \frac{y_g}{x_g}-\frac{y_{h}}{x_{h}} \right).
\end{align}
Notice that $y_g/x_g$ is the contribution of the group $g$ to the slope of 
the prediction curve in the hit$\times$false-alarm space.
If the $G$ groups are ordered by decreasing $P(E|g)$, then $y_g/x_g$ also 
decreases with $g$.
Therefore $(y_g/x_g - y_h/x_h)>0$ for $g<h$ and $(y_g/x_g - y_h/x_h)>0$ 
for $g>h$ and Eq.~(\ref{eq.y}) is maximized by choosing $\pi_g$ such that the 
two last terms vanish.
This is achieved choosing
\begin{equation}\label{eq.pig}
  \pi_g = \begin{cases}
      1 & g<h, \\
      \frac{x-\sum_{g< h} x_g}{x_h} & g=h, \\
      0 & g>h, \\
          \end{cases}
\end{equation}
which correspond to issuing positive predictions only to the $h$ groups with 
largest \footnote{Positive events are predicted for the group $h$ in
  Eq.~(\ref{eq.pig}) as much as needed to reach the required false positive 
rate 
  $x$.} 
$P(E|g)$ and is equivalent to strategy (LD) mentioned in the main text.

\subsection{Computation of $\Pi$ for the optimal strategy}

As illustrated in Fig.~\ref{fig.2}(b), the partition performed by the 
optimal strategy defines $G$ different intervals in the hit and false alarm axis
(the points for which $P(E|g)=P_*$, $g\in\{1\dots G\}$) and therefore $G^2$ 
rectangles in the hit$\times$false-alarm space.
The $(g,h)$ rectangle has height $P(h)P(E|h)/P(E)=P(h|E)$, width
$P(g|\bar{E})$ (where $\bar{E}$ is the complement of $E$, i.e., 
$P(\bar{E}|g)= 1-P(E|g)$), and therefore it has an area 
$A_{g,h}= P(h|E)P(g|\bar{E})$.
The curve of strategy (LD) is the union of the diagonals of the $g=h$ 
rectangles (which are obtained by increasing $p_*$).
$\Pi$ is two times the sum of the rectangles and triangles under this 
curve minus half of all the area: 
\begin{equation}
  \begin{array}{ll}
  \Pi & = 2\left[ \sum_g \sum_{h<g} A_{g,h} +\dfrac{1}{2} \sum_g A_{g,g} 
-\dfrac{1}{2}
  \sum_g \sum_h A_{g,h}\right] \\ \\
    & = \sum_g \sum_{h<g} A_{g,h} - \sum_g \sum_{h>g} A_{g,h} \\ \\
    & = \sum_g \sum_{h<g} (A_{g,h} - A_{h,g}) \\ \\
    & = \sum_{g}\sum_{h<g} P(h|E)P(g|\bar{E})-P(h|\bar{E})P(g|E)\\ \\
    & 
=\dfrac{\sum_{g}\sum_{h<g}P(g)P(h)\left(P(E|h)-P(E|g)\right)}{P(E)(1-P(E))},
%   \label{eq.auc2}
  \end{array}
\end{equation}
where we used $\sum_g \sum_h A_{g,h}=1$. This finishes our demonstration of 
Eq.~(\ref{eq.pi}).

\section*{Acknowledgments}
We thank M. Gerlach, S. Hallerberg, and S. Siegert for insightful discussions 
and M. Gerlach and S. Bialonski for careful reading of the manuscript.

\bibliography{preprint}

% \section*{Figure Legends}

% \section*{Tables}

\end{document}